\documentclass[prl,superscriptaddress,tightenlines,twocolumn,balancelastpage]{revtex4}
\usepackage{amsfonts}
\usepackage{amsmath}
\usepackage{amssymb}
\usepackage{graphicx}
\usepackage{dcolumn}

\begin{document}

\preprint{}
\title[IR of Bi2Te3]{Magneto-Infrared Spectroscopic Study of Thin Bi$_{2}$Te$
_{3}$ Single Crystals}
\author{L.-C. Tung}
\affiliation{National High Magnetic Field Laboratory, Tallahassee, Florida 32310}
\author{W. Yu}
\affiliation{School of Physics, Georgia Institute of Technology, Atlanta, Georgia, 30332}
\author{P. Cadden-Zimansky}
\affiliation{Physics Program, Bard College, Annandale-on-Hudson, New York 12504}
\author{I. Miotkowski}
\affiliation{Department of Physics, Purdue University, West Lafayette, Indiana 47907}
\author{Y. P. Chen}
\affiliation{Department of Physics, Purdue University, West Lafayette, Indiana 47907}
\author{D. Smirnov}
\affiliation{National High Magnetic Field Laboratory, Tallahassee, Florida 32310}
\author{Z. Jiang}
\affiliation{School of Physics, Georgia Institute of Technology, Atlanta, Georgia, 30332}
\date{\today }

\begin{abstract}
Thin Bi$_{2}$Te$_{3}$ single crystals laid on Scotch tape are investigated
by Fourier transform infrared spectroscopy at $4$K and in a magnetic field
up to $35$T. The magneto-transmittance spectra of the Bi$_{2}$Te$_{3}$/tape
composite are analyzed as a stacked-slab system, and the average thickness
of Bi$_{2}$Te$_{3}$ is estimated to be $6.4\pm 1.7\mu $m. The optical
conductivity of Bi$_{2}$Te$_{3}$ at different magnetic fields is then
extracted, and we find that magnetic field modifies the optical conductivity
in the following ways: (1) Field-induced transfer of the optical weight from
the lower frequency regime ($<250$cm$^{-1}$) to the higher frequency regime (%
$>250$cm$^{-1}$) due to the redistribution of charge carriers across the
Fermi surface. (2) Evolving of a Fano-resonance-like spectral feature from
an anti-resonance to a resonance with increasing magnetic field. Such
behavior can be attributed to the electron-phonon interactions between the $%
E_{u}^{1}$ optical phonon mode and the continuum of electronic transitions.
(3) Cyclotron resonance resulting from the inter-valence band Landau level
transitions, which can be described by the electrodynamics of massive Dirac
holes.
\end{abstract}

\volumeyear{year}
\volumenumber{number}
\issuenumber{number}
\eid{identifier}
\received[Received text]{date}
\revised[Revised text]{date}
\accepted[Accepted text]{date}
\published[Published text]{date}
\startpage{101}
\endpage{102}
\maketitle

Recently, bismuth telluride (Bi$_{2}$Te$_{3}$), along with Bi$_{2}$Se$_{3}$,
Sb$_{2}$Te$_{3}$, and other V$_{2}$VI$_{3}$ binary alloys, were discovered
to be a physical realization of a three-dimensional (3D) topological insulator (TI).
\cite{Rev} The insulating bulk of a TI is enclosed by a robust conducting surface,
which exhibits a linear Dirac-like band structure protected by
time-reversal symmetry. The strong spin-orbit coupling in TIs also leads to
unusual spin-momentum locking\cite{Hsi09} as well as unique magnetoelectric
effects.\cite{Tse10} These phenomena may have profound implications in future
nanoelectronic, spintronic and thermoelectric devices.\cite{Plu02,Fu08,Moo10}
More TIs are discovered and/or proposed, including SrTiO$_{3}$/LaAlO$_{3}$
interface\cite{Oht04,Rey07}, InAs/GaSb coupled quantum well\cite{Kne11},
HgTe under strain\cite{Cle11}, grey tin\cite{Fu07}, and topological oxides.%
\cite{Shi09,Hos13,Yan14}

Since its discovery, considerable efforts have been made to understand the
electronic properties of both surface and bulk charge carriers in 3D TIs.
These properties can be studied via optical techniques including
angle-resolved photoemission spectroscopy, optical conductivity, and
magneto-infrared spectroscopy. The optical conductivity and cyclotron
resonance (CR) of several TI materials have been reported.\cite{Kul99,Ste07,
But10,Laf10,Sch12,Pie12,Val12,Cha14,Wu15,Orl15} In particular, transitions between the
quantized Landau levels (LLs) of the Dirac surface states have been observed
in Bi$_{0.91}$Sb$_{0.09}$, though it seems to suggest three distinct Fermi
velocities, thus implying three Dirac cones on the [$2\overline{1}\overline{1%
}$] surface.\cite{Sch12} In Bi$_{2}$Se$_{3}$, interband LL transitions have
also been observed and they can be attributed to the electrodynamics of
massive Dirac fermions in TIs.\cite{Orl15} In this Report, magneto-infrared
transmittance spectroscopy is used to obtain the optical
conductivity of thin Bi$_{2}$Te$_{3}$ single crystal flakes laid on a strip
of Scotch tape. The thickness of the Bi$_{2}$Te$_{3}$ is reduced by
exfoliation, in order to suppress the bulk contribution and allow for
appreciable infrared transmission. Transmittance spectroscopy is preferred
for resolving optical modes in TIs, because an optical mode generally
results in an absorption dip in the transmittance spectrum. Our measurements
reveal that a Fano resonance\cite{Fan61} occurs at the $E_{u}^{1}$ optical
phonon mode\cite{Ric77,Kul84} which evolves from an anti-resonance
(enhanced transmission) to a resonance (absorption) with increasing magnetic
field. In addition, a broad magnetic-field dependent absorption is observed
due to the inter-valence band LL transitions of massive Dirac holes.\cite%
{Cha14,Orl15,Wol64,Pal70}

Bi$_{2}$Te$_{3}$\ crystallizes in a rhombohedral structure (point group $%
\overline{3}mD_{3}d$) with quintuple layers stacked along the trigonal $c$%
-axis.\cite{Bla57} The neighboring quintuple layers are bounded by weak van
der Waals forces, allowing for exfoliation of Bi$_{2}$Te$_{3}$ thin layers.
In this work, we first exfoliate a thin Bi$_{2}$Te$_{3}$ crystal from a
larger piece and lay it on a strip of Scotch tape. The thickness of this
thin crystal is further reduced by repeatedly lifting layers from it.
Optical microscopy investigation reveals that the resulting sample consists
of thousands of thin flakes\ with most of them exhibiting an area of $%
100-300\mu $m$^{2}$. The Bi$_{2}$Te$_{3}$/tape composite is then placed in a
liquid helium cryostat held at $4$K subject to a high magnetic field up to $%
35$T. Magneto-infrared transmission spectra are measured by a Fourier
transform infrared spectrometer using light pipe optics, and the
transmittance is obtained by the ratio to a reference spectrum taken in the
absence of the sample.

\begin{figure}[tp]
{{{%
\includegraphics[
]{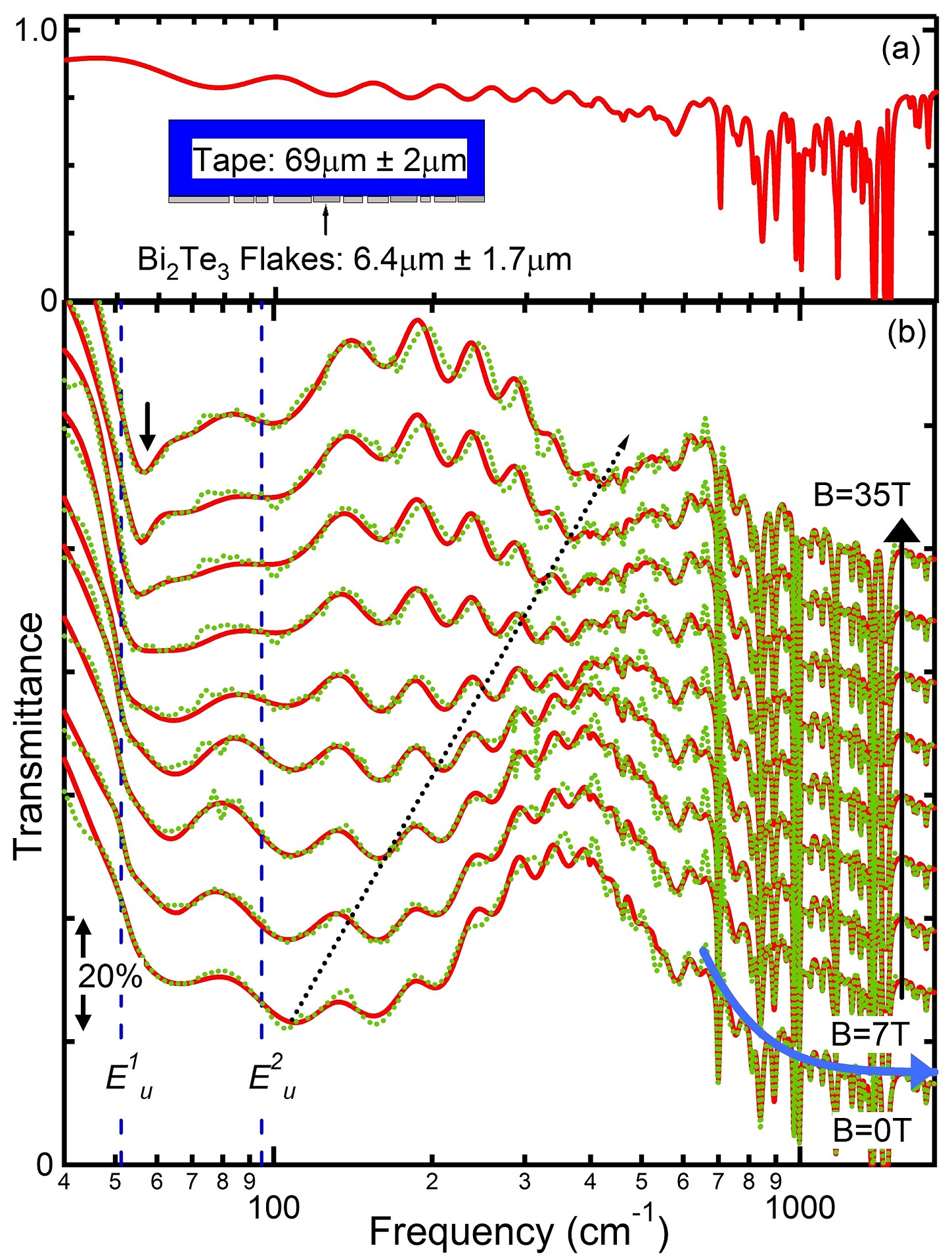}}}}
\caption{(Color online) (a) The transmittance spectrum of the Scotch tape.
The tape is transparent at low frequencies ($<700$cm%
$^{-1}$), while sharp absorption lines appear at $>700$cm$^{-1}$. 
These sharp lines have minimal effects to the spectral features discussed 
in this work, as those features are much broader and occur at lower frequencies.
Inset: The schematic layout of the Bi$_{2}$Te$_{3}$ flakes on Scotch tape. (b) The 
magneto-transmittance spectra (green) of the Bi$_{2}$Te$_{3}$/tape
composite. The spectra are shown in every $4$T from $7$T to $35$T and each
spectrum is offset vertically for clarity. Red solid lines represent the
best fits to the data via modeling the composite as a stacked--slab system
and considering parallel interface interference. The (blue) arrow
approaching $1200$cm$^{-1}$ indicates where the transmittance drops rapidly with
increasing photon frequency. The short arrow around $50$cm$^{-1}$ labels the
spectral feature that can be attributed to a Fano resonance. A broad dip
gradually shifts to higher frequencies with increasing magnetic field as
indicated by the dotted arrow. The infrared-active optical phonon modes are
marked with (blue) dashed lines with the values taken from Ref. \protect\cite{Ric77}.}
\end{figure}

A set of magneto-transmittance spectra of the Bi$_{2}$Te$_{3}$/tape
composite at different magnetic fields is shown in Fig. 1(b) and compared
with that of the tape at zero field in Fig. 1(a). Here, as one can see, the
Bi$_{2}$Te$_{3}$/tape composite is transparent at low frequencies, while the
transmittance drops rapidly approaching $1200$cm$^{-1}$ ($1\text{eV}\simeq8065$cm%
$^{-1}$) as indicated by the (blue) arrow in Fig. 1(b). This drop in transmittance
occurs in all the spectra taken at different magnetic fields, and it can be attributed to
the rise of the conductance/absorption due to the electronic transitions from
the valence to conduction band. The drop-off frequency ($\sim$1200cm$^{-1}$)
implies a bandgap around $0.15$eV, consistent with that reported in
literature.\cite{Book,Kul94} The parallel surfaces of the tape and the Bi$%
_{2}$Te$_{3}$ flakes cause a series of etalon oscillations. The period of the
oscillation depends on the magnetic field owing to the field-dependent refractive
index of Bi$_{2}$Te$_{3}$. Nevertheless, a discernible dip located around 
$50$cm$^{-1}$ (as indicated by the short arrow in Fig. 1(b))
develops at high magnetic fields, while another broader dip gradually blue
shifts to higher frequencies as guided by the dotted arrow.

To extract the optical conductivity of Bi$_{2}$Te$_{3}$, the acquired
spectra of the Bi$_{2}$Te$_{3}$/tape composite\ are modeled as a
stacked-slab system considering the etalon oscillations due to the parallel
interfaces of each slab using RefFit.\cite{Kuz05} The dielectric function of each layer
is modeled by a set of Drude-Lorentz modes as $\varepsilon _{1}(\omega
)+i\varepsilon _{2}(\omega )=\varepsilon _{\infty }+\sum_{j}\frac{\omega
_{pj}^{2}}{\omega _{0j}^{2}-\omega ^{2}-i\gamma _{j}\omega }$, where $%
\varepsilon _{\infty }$ is the high-frequency dielectric constant. The
parameters $\omega _{pj}$, $\omega _{0j}$, and $\gamma _{j}$ are the plasma
frequency, the oscillator's natural frequency, and the linewidth of the $j$%
-th Lorentz oscillator, respectively. The real part of the optical
conductivity can then be calculated by $\sigma _{1}=\frac{\omega \varepsilon
_{2}}{4\pi }$. It is important to note that the net transmittance spectra of
the Bi$_{2}$Te$_{3}$ flakes cannot be directly obtained by the ratio of the
transmittance spectra of the Bi$_{2}$Te$_{3}$/tape composite to that of the
tape, because the tape-vacuum and tape-Bi$_{2}$Te$_{3}$ interfaces have very
different Fresnel coefficients. The best fits to the data are shown in Fig. 1(b) for 
each magnetic field. An asymmetric, Fano-resonance-like spectral feature is 
revealed at low frequencies near the infrared-active $E_{u}^{1}$ optical phonon
mode, which will be discussed below in the context of Fig. 2(b).

From the transmittance spectrum of the tape (Fig. 1(a)), the dielectric
function and the thickness of the tape can be determined from the etalon
oscillations. These parameters are held constant when analyzing the
magneto-transmittance spectra of the Bi$_{2}$Te$_{3}$/tape composite, and
the best fits to the data in Fig. 1(b) (red solid lines) are obtained using $%
69\pm 2\mu $m for the thickness of the tape and $6.4\pm 1.7\mu $m as the
average thickness of Bi$_{2}$Te$_{3}$ flakes. For the sake of simplicity, in
this work we regard the flakes as a uniform thin slab, disregarding the
existence of the surface layers. The optical conductivity of Bi$_{2}$Te$_{3}$
can then be calculated from the best fit of the magneto-transmittance spectrum. 
\begin{figure}[tp]
{{{%
\includegraphics[
]{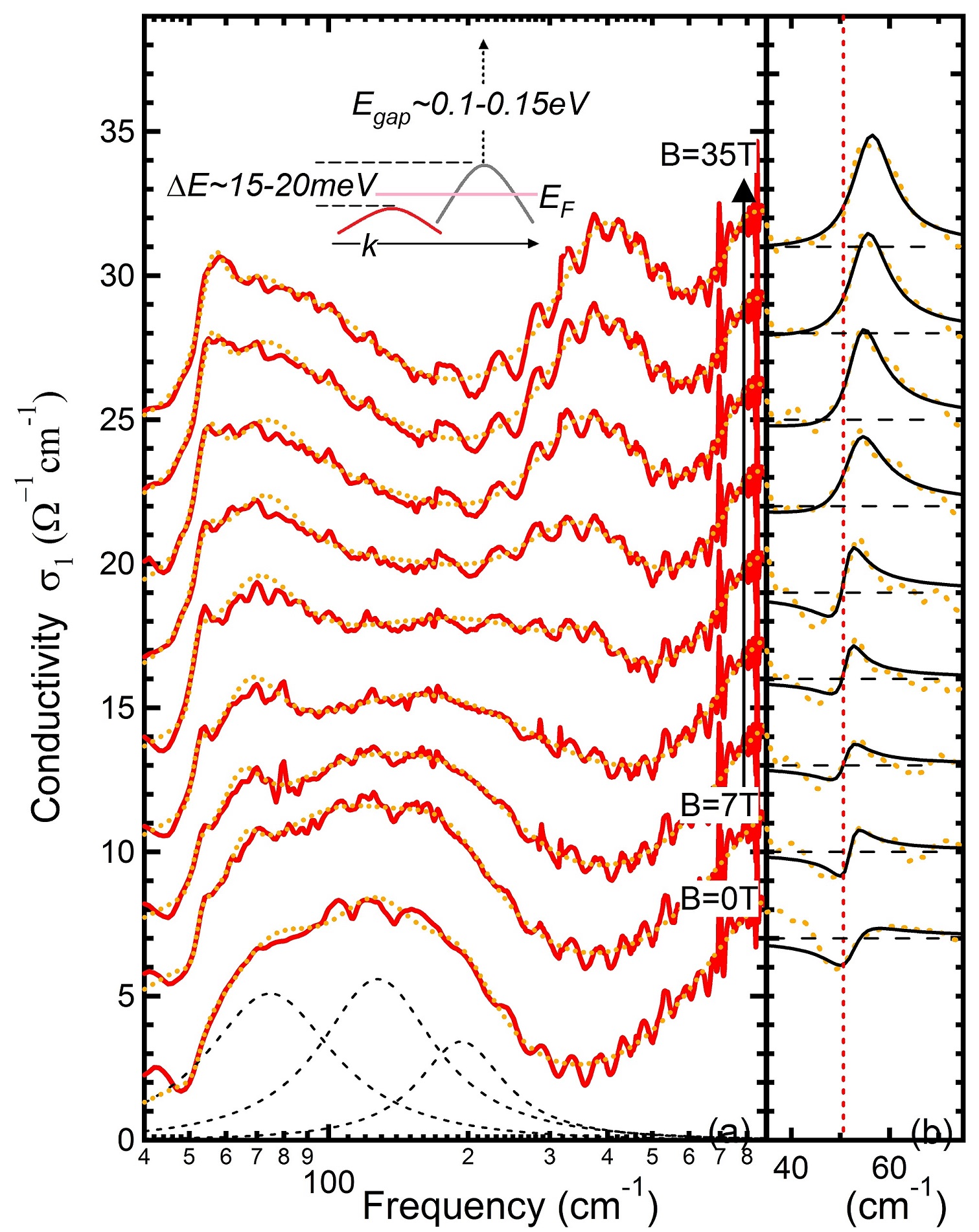}}}}
\caption{(Color online) (a) The real part of the optical conductivity of Bi$%
_{2}$Te$_{3}$. The spectra are shown in every $4$T from $7$T to $35$T and
each spectrum is offset vertically for clarity. The (orange) dashed lines represent
the as-calculated optical conductivity from the best fits to the data, and the (red)
solid lines represent the corrected optical conductivity which considers the differences
between the best fits and the measured transmittance spectra in Fig. 1(b).
Three Lorentzian modes, shown in (black) dashed lines, are used to describe
the broad spectral feature centered at $\sim$150cm$^{-1}$. Inset: Schematic
band structure of Bi$_{2}$Te$_{3}$ reproduced from Ref. \protect\cite{Kul94}%
. (b) The Fano resonance is separated for further analysis. The solid lines
represent the best fits to the data using Fano formula.}
\end{figure}

The optical conductivity spectra of Bi$_{2}$Te$_{3}$ at selected fields are
plotted in Fig. 2 (a). To ensure that minute or non-Lorentzian conductivity
contributions are not neglected by modeling, the difference between the fits
and the measured transmittance spectra in Fig. 1 (b) are calculated, and a
correction is applied to the as-calculated conductivity spectra by adding a
conductivity correction $\Delta \sigma _{1}=\frac{n_{1}c}{4\pi \mu _{1}}%
\Delta \alpha \simeq \frac{n_{1}c}{4\pi \mu _{1}d}\frac{\Delta T}{T}$, where 
$\mu _{1}$ is the magnetic permeability which is assumed to be $1$ for
non-magnetic materials, $n_{1}$ is the real part of the refractive index
which is obtained from the dielectric function, $d$ is the average thickness
of the flakes, $c$ is the speed of light, and $\Delta \alpha $\ represents a
correction to the absorption coefficient which amounts to the observed
transmittance difference. In Fig. 2 (a), we plot the corrected conductivity
spectra overlaid on top of the as-calculated spectra for comparison. It is
clear that the model did not completely remove the etalon oscillations due to
its over-simplicity. Nevertheless, the period of the oscillations in the
corrected conductivity spectra agrees with the thickness of the tape,
extracted from Fig. 1(a).

At low magnetic fields, a broad conductivity peak (modeled as three
Lorentzian modes in Fig. 2(a)) is evidenced at $\sim$150cm$^{-1}$. We notice that
a double-Lorentzian spectral structure around $200$cm$^{-1}$\ was previously 
observed in Bi$_{2}$Te$_{2}$Se and attributed to the transitions from the impurity states to the electronic
continuum states.\cite{Pie12} In this study, Bi$_{2}$Te$_{3}$ is $p$-doped%
\cite{Haj14} and the center of this broad peak coincides with the band
offset between the upper and the lower valence bands.\cite{Kul94} Therefore,
we attribute it to the continuum of the electronic transitions from
the lower valence band to the empty states above the Fermi surface in the
upper valence band.

The conductivity mode located at $\sim$50cm$^{-1}$ invokes
the asymmetric lineshape of a Fano resonance\cite{Fan61} and consistently
evolves from an anti-resonance (a dip) to a resonance (a peak) with
increasing magnetic field as shown in Fig. 2(b). Fano resonance is
ubiquitous across several branches of physics, and it describes the quantum
interference between two coupled transition pathways: one via a discrete
excited state and the other via a continuum of states. In this study, the
discrete state is the $E_{u}^{1}$ optical phonon mode\cite{Ric77,Kul84},
while the continuum of states are the transitions from the lower valence band
to the empty states in the upper valence band. In the presence of an applied
magnetic field, the inter-valence band transitions are dominated by the
quasi-continuous Landau level transitions with linewidth larger than $%
70 $cm$^{-1}$. The optical phonon mode and the continuum of the electronic
transitions are coupled via strong electron-phonon interactions, which can be attributed to the
"charged-phonon theory"\cite{Ric92,Cap12} and/or the topological
magnetoelectric effect.\cite{Laf10} A magnetic-field tunable Fano resonance
was also found in Bi$_{2}$Se$_{3}$ ($\sim$64cm$^{-1}$; $\alpha$-phonon
mode), and attributed to the topological magnetoelectric and
magnetostriction effects.\cite{Laf10} Various types of Fano resonance
resulting from the coupling between the optical phonon modes and
electronic transitions were observed in Bi$_{2}$Se$_{2}$Te and graphitic
materials.\cite{Pie12,Cap12}

To separate the optical conductivity of the Fano resonance for quantitative
analysis, other Lorentzian modes are removed and the contribution of the Fano
resonance is isolated and plotted in Fig. 2(b). The conductivity of a Fano
resonance can be described by $\Delta \sigma _{Fano}=\sigma _{0,Fano}\frac{%
q^{2}+2qz-1}{q^{2}(1+z^{2})}$ with $z=\frac{\omega -\omega _{Eu}}{\gamma }$
and $\sigma _{0,Fano}=\frac{\omega _{p}^{2}}{4\pi \gamma }$, where $\omega
_{p}$ is the plasma frequency, $\omega _{Eu}$ is the phonon energy, $\gamma $
is the linewidth, and $q$ is the dimensionless Fano parameter which
characterizes the resonance lineshape. When $\left\vert q\right\vert \ll 1$,
the Fano effect results in an anti-resonance with asymmetric lineshape. When 
$\left\vert q\right\vert \gg 1$, it becomes a resonance and its lineshape
can be well approximated by a Lorentzian mode.

\begin{figure}[tp]
{{{%
\includegraphics[
]{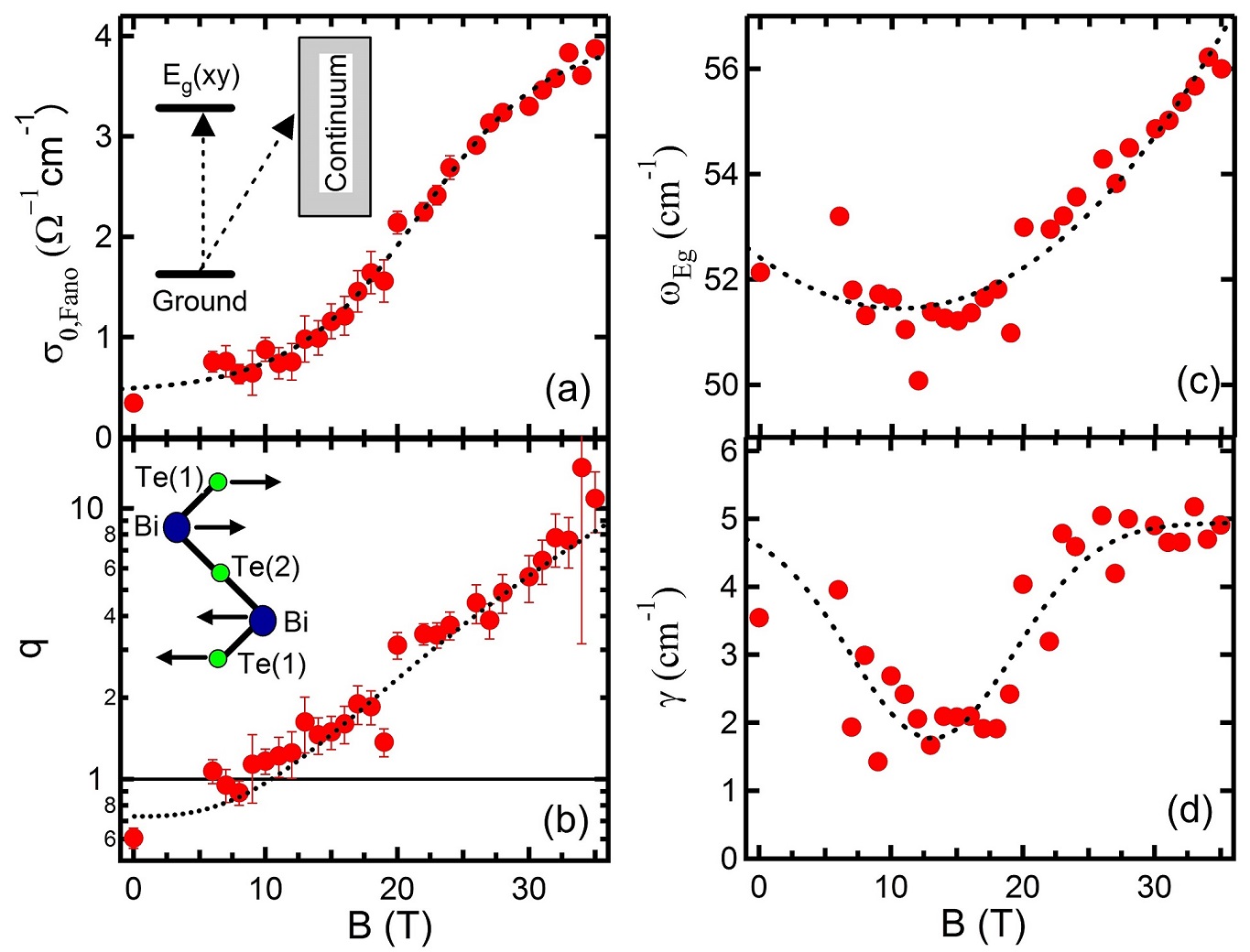}}}}
\caption{(Color online) (a) $\protect\sigma _{0,Fano}$ as a function of the
magnetic field. Inset: Schematic energy level diagram of the Fano resonance.
(b) The dimensionless Fano parameter $q$ as a function of the magnetic
field. Inset: The displacement of ions of the $E_{u}^{1}$ phonon mode. (c)
The energy of the optical phonon mode $E_{u}^{1}$ as a function of the
magnetic field. (d) The linewidth of the Fano resonance $\protect\gamma $\
as a function of the magnetic field. The dotted lines are drawn as guides for 
the magnetic-field dependence; they do not result from a theoretical model.}
\end{figure}

As shown in Fig. 2(b), the low-frequency conductivity contribution can be well fitted by
the asymmetric lineshape of a Fano resonance. The fitting parameters at
different magnetic fields are plotted in Fig. 3. In Fig. 3(b), the Fano
parameter $q$ gradually increases from $q<1$ to $q>>1$ with increasing
magnetic field crossing over $q\simeq 1$\ around $B\simeq 10$T. Similarly,
the optical strength $\sigma _{0,Fano}$\ of the Fano resonance and the $%
E_{u}^{1}$ phonon energy start to rise at the crossover field as shown in
Fig. 3(a) and (c), respectively. The linewidth $\gamma $\ appears to find a
minimum at $B\simeq 10$T as shown in Fig. 3(d).

The observed Fano resonance, as well as that reported in Bi$_{2}$Se$_{3}$%
, is unusual, because the magnetic-field induced modification of the Fano
effect is rarely found in non-magnetic systems.\cite{Laf10} In addition, it
involves an interaction that couples a lattice excitation (an optical phonon
mode) and the electronic transitions between the valence bands. The
electron-phonon interactions between the $\alpha$-phonon mode and the
electronic transitions in Bi$_{2}$Se$_{3}$ were attributed to the
magnetostriction in the system with enhanced spin-orbit coupling and the
magnetoelectric coupling in a non-trivial TI.\cite{Laf10} A local electric
field, induced by the applied magnetic field via the magnetoelectric effect%
\cite{Ess09}, modifies the motion of the Bi ions and thus the optical phonon
mode.\cite{Laf10} In Bi$_{2}$Se$_{3}$, the Fano resonance is observed when
the magnetic field is applied in the direction parallel to the displacement
of the Bi ions in the $\alpha$-phonon mode\cite{Laf10}, whereas in Bi$_{2}$%
Te$_{3}$, it is observed near the $E_{u}^{1}$ optical phonon mode in which the
displacement of the Bi ions are perpendicular to the magnetic field 
direction\cite{Ric77,Kul84}, as shown in the inset to Fig. 3(b).

The observed Fano-resonance-like behavior is also consistent with the
``charged phonon theory"\cite{Ric92,Cap12}, in which the optical phonon mode
``borrows" optical strength (charges) from the electronic transitions or
vice versa, depending on the relative energies of the electronic and the
optical transitions. This theory has been successfully adopted to explain
the Fano effect in graphitic materials.\cite{Cap12} In our case, the
electronic transitions gradually shift to higher energies due to the
diamagnetic shift of the Fermi level and at $B\simeq 10$T it reaches the energy of the $E_{u}^{1}$
phonon mode. For $B>10$T, the optical phonon mode borrows the charges from
the electronic transitions, thus resulting in the rise of the optical
strength and the anti-resonance to resonance crossover. The crossover field
is consistent with the expected diamagnetic shift of the Fermi level with a
hole mass equal to $0.08m_{e}$\cite{Book}, where $m_{e}$ is the free
electron mass.

Other than the Fano resonance, the applied magnetic field gradually
transfers the optical weight from the broad peak centered at $\simeq$150cm$%
^{-1}$ to higher frequencies. Magnetic field is expected to induce the
transitions between quantized LLs between the lower and the upper
valence bands. The degeneracy and the separation of the LLs increase with
increasing magnetic field, resulting in the observed optical weight
transfer and the increase of the transition energy. The magnetic-field
induced effect can be evaluated by subtracting the optical conductivity obtained
at $B=0$T from that obtained at $B\neq 0$T: $\Delta \sigma _{B}=\sigma
_{B}(\omega )-\sigma _{B=0T}(\omega )$. The magnetic-field induced change in
optical conductivity is plotted in Fig. 4(c), where a broad peak develops as increasing
magnetic field and the peak gradually shifts to higher frequencies with
increasing optical strength ($\varpropto $ the enclosed area). 
\begin{figure}[tp]
{{{%
\includegraphics[
]{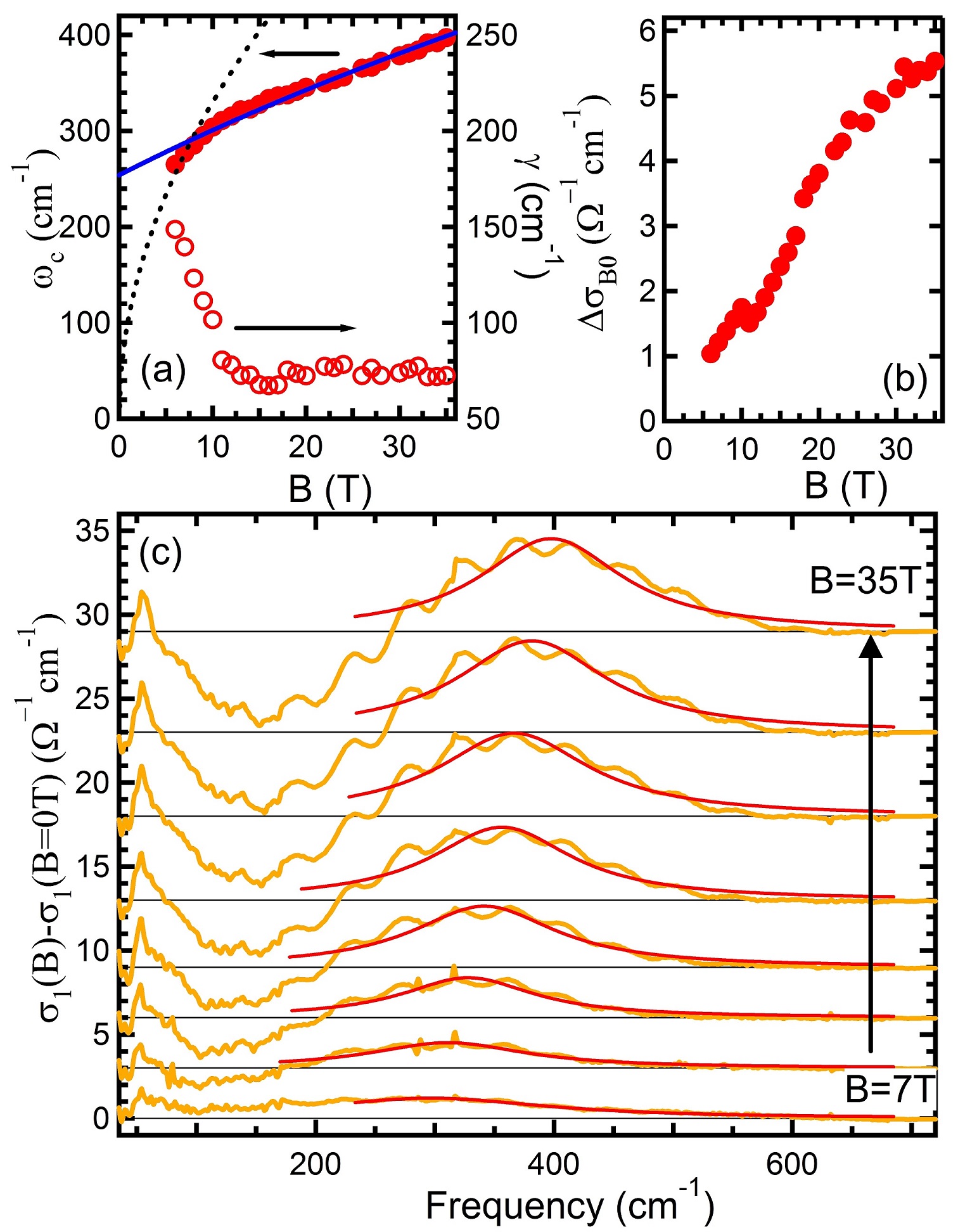}}}}
\caption{(Color online) (a) The transition energy $\protect\omega _{c}$ and
the linewidth $\protect\gamma $\ as a function of the magnetic field. The
(blue) solid line represents the best fitting curve to Eq. (1). The (black)
dashed line represents the transition energy from the $n=0$ LL to the $n=1$
LL in the Dirac surface state with a Fermi velocity $v_{F}\simeq 0.36\times
10^{6}m/s$. (b) The optical strength $\Delta \protect\sigma _{B0}$\ as a
function of the magnetic field. (c) The magnetic-field induced optical
conductivity change $\Delta \protect\sigma _{B}$ from 7T to 35T with 4T 
interval. Each conductivity spectrum is offset vertically for clarity. The
horizontal lines label the positions where the change of the optical
conductivity is zero and the red solid lines are the Lorentzian fits to the data.}
\end{figure}

The magnetic-field induced broad peak is analyzed as a CR resulting from LL
transitions using $\Delta \sigma _{B}=\frac{\Delta \sigma
_{B0}}{1+(\frac{\omega -\omega _{c}}{\gamma })^{2}},$ where the amplitude $%
\Delta \sigma _{B0}=\frac{\omega _{p}^{2}}{4\pi \gamma }$ represents the
optical strength of the CR, $\omega _{c}$ is the cyclotron frequency, and $%
\gamma $ is the linewidth. The cyclotron frequency $\omega _{c}$ and the
linewidth $\gamma$ of the CR are plotted in Fig. 4(a), while the optical strength 
$\Delta \sigma _{B0}$ is plotted in Fig. 4(b). The optical strength rises with
magnetic field due to the increasing LL degeneracy and density of states 
in a 3D electron system. It is clear that the transition energy does not extrapolate 
to zero at zero magnetic field, so the observed CR is not a consequence of the 
intraband LL transitions. Moreover, the energy of the CR falls in the range of 
several tens of meV, whereas the excitonic interband LL transitions (from the valence
to conduction bands) reside in the range of several hundred meV.\cite{Orl15} We
notice that the observed transition energy is close to the band offset between the 
lower and upper valence bands in Bi$_2$Te$_3$ (Inset to Fig. 2(a)). Therefore, 
the observed magnetic-field dependent mode can be attributed to the transitions 
from the LLs in the lower valence band to that in the upper valence band.

Next, we will attempt to quantitatively describe the magnetic-field
dependence of $\omega _{c}$ in Fig. 4(a) as an inter-valence band LL
transition. The holes in the bulk valence bands are described as massive
Dirac fermions, and the energies of the LLs of massive Dirac fermion
disperse as\cite{Wol64,Pal70}

\begin{equation}
E=\pm \sqrt{\frac{E_{g}{}^{2}}{4}+E_{g}\left[ \frac{\hbar eB}{2m^{\ast }}(n+%
\frac{1}{2})\pm \frac{1}{2}g_{0}\mu _{B}B\right] },
\end{equation}%
where $E_{g}$ is the energy of the direct bandgap, $\mu _{B}$ is the Bohr
magneton, $m^{\ast }$ is the effective mass of the carriers, $n$ is the LL
index, and $g_{0}$ is Land\'{e} g-factor. The first $\pm $ sign indicates
the energy of the conduction (+) or the valence (-) band, while the second $%
\pm $ sign sets the spin state of the LL. For simplification, we assume that
the LLs of both valence bands can be described by Eq. (1), but with
different bandgap and effective mass, disregarding their relative
positions in the Brillouin zone (as shown in the inset to Fig. 2(a)). We
also neglect the spin-flip transitions and set the same g-factor for both
bands. The transition energy is then insensitive to the choice of the $g_{0}$
value. The hole effective mass $m_{h,up}^{\ast }$ of the upper valence band
is selected to be $0.08m_{e}$, because this value was also determined by a
CR measurement.\cite{Book} The bandgap of the upper valence band is given as 
$E_{g,up}\simeq 0.15$eV.\cite{Kul94} The effective mass $m_{h,low}^{\ast }$
and the bandgap $E_{g,low}$\ of the lower valence band are left as fitting
parameters. The transition energy $\omega _{c}$\ can then be attributed to
the transition from the $n=1$ LL in the lower valence band to the $n=0$ LL
in the upper valence band. The calculated magnetic-field dependence of the
transition energy is plotted in Fig. 4(a) in a blue solid line. From the
best fitting curve, the hole effective mass of the lower valence band is
found to be $m_{h,low}^{\ast }\simeq 0.088m_{e}$ and the bandgap 
$E_{g,low}\simeq 0.21$eV. The hole effective mass of the lower valence band is
very close to the reported value of $m^{\ast }=0.09m_{e}$ for the second
hole band\cite{Book}, although the band offset between the valence bands
is found to be $\Delta E\simeq 30$meV from this simple model, larger than
$\Delta E\simeq 20$meV estimated from transport
measurements.\cite{Koh76} From these band parameters, one can evaluate the
band velocity $v_{D}$\ of the massive Dirac fermions by $v_{D}=\sqrt{%
E_{g}/m^{\ast }}$. The band velocities are $v_{D}\simeq 0.58\times 10^{6}m/s$
and $v_{D}\simeq 0.65\times 10^{6}m/s$ for the upper and lower valence bands,
respectively. These values are slightly larger than that (%
$v_{D}\simeq 0.46\times 10^{6}m/s$) found in Bi$_{2}$Se$_{3}$.\cite{Orl15}

Transitions between the LLs of the Dirac surface states cannot be
distinguished from bulk LL transitions due to the broad linewidth of the CR.
However, we notice that the transition energy $\omega _{c}$ deviates from
Eq. (1) at $B<10$T, while the linewidth becomes substantially broader (Fig.
4(a)). The dashed line in Fig. 4(a) illustrates the contribution of the $%
n=0\rightarrow 1$ LL transition of the Dirac surface states with a Fermi
velocity $v_{F}\simeq 0.36\times 10^{6}m/s$.\cite{Haj12} It is possible that
the CR of the surface carriers indeed contributes to the optical
conductivity at low magnetic fields. At a higher field, however, the
diamagnetic shift of the Fermi level depletes the carriers in the surface
band, and thus the inter-valence band LL transitions dominate.

In summary, the optical conductivity of thin Bi$_{2}$Te$_{3}$ single
crystals is studied by the magneto-infrared transmittance spectroscopy in
high magnetic fields. An anomalous field-tunable Fano resonance is observed,
and it is attributed to the electron-phonon interactions between the $E_{u}^{1}$
optical phonon mode and the continuum of the electronic transitions in
the valence bands. The strong electron-phonon interactions might
be suggestive of the topological magnetoelectric effect. In addition, a broad
absorption mode is found to increase in the optical strength and the central
energy with increasing magnetic field, and it can be attributed to the
transition from the $n=1$ LL in the lower valence band to the $n=0$ LL in
the upper valence band. The CR of the surface LLs cannot be clearly
identified in our measurement, but it might contribute to the unusual
broadening of the absorption at lower fields ($B<10$T). Further experimental
studies are needed on intrinsic TI materials with a truly insulating bulk.

\begin{acknowledgments}
This work is supported by the DOE (Grant No. DE-FG02-07ER46451). The TI crystal
synthesis at Purdue University is supported by the DARPA MESO program
(Grant No. N66001-11-1-4107). The infrared measurements are carried out at the
National High Magnetic Field Laboratory, which is supported by NSF Cooperative
Agreement No. DMR-0654118, by the State of Florida, and by the DOE.
\end{acknowledgments}


\begin{thebibliography}{99}
\bibitem{Rev} For recent reviews, see for example, M. Z. Hasan and C. L. Kane,
Rev. Mod. Phys. \textbf{82}, 3045 (2010) and X.-L. Qi and S.-C. Zhang, Rev.
Mod. Phys. \textbf{83}, 1057 (2011).

\bibitem{Hsi09} D. Hsieh, Y. Xia, D. Qian, L. Wray, J. H. Dil, F. Meier, J.
Osterwalder, L. Patthey, J. G. Checkelsky, N. P. Ong, A. V. Fedorov, H. Lin, A.
Bansil, D. Grauer, Y. S. Hor, R. J. Cava, and M. Z. Hasan, Nature \textbf{460},
1101 (2009).

\bibitem{Tse10} W.-K. Tse and A. H. MacDonald, Phys. Rev. B \textbf{82},
161104 (2010).

\bibitem{Plu02} K. J. Pluciski, W. Gruhn, I. V. Kityk, W. Imioek, H.
Kaddouri, and S. Benet, Optics Communications \textbf{204}, 355
(2008).

\bibitem{Fu08} L. Fu and C. L. Kane, Phys. Rev. Lett. \textbf{100}, 096407 (2008).

\bibitem{Moo10} J. E. Moore, Nature \textbf{464}, 194 (2010).

\bibitem{Oht04} A. Ohtomo and H. Y. Hwang, Nature \textbf{427}, 423 (2004).

\bibitem{Rey07} N. Reyren, S. Thiel, A. D. Caviglia, L. Fitting Kourkoutis,
G. Hammerl, C. Richter, C. W. Schneider, T. Kopp, A.-S. R\"{u}etschi, D.
Jaccard, M. Gabay, D. A. Muller, J.-M. Triscone, and J. Mannhart, Science 
\textbf{317}, 1196 (2007).

\bibitem{Kne11} I. Knez, R. R. Du, and G. Sullivan, Phys. Rev. B \textbf{81}, 
201301(R) (2010).

\bibitem{Cle11} C. Bouvier, T. Meunier, R. Kramer, and L. P. Levy, arXiv:1112.2092.

\bibitem{Fu07} L. Fu and C. L. Kane, Phys. Rev. B \textbf{76}, 045302 (2007).

\bibitem{Shi09} A. Shitade, H. Katsura, J. Kune\v{s}, X.-L. Qi, S.-C. Zhang,
and N. Nagaosa, Phys. Rev. Lett. \textbf{102}, 256403 (2009).

\bibitem{Hos13} J. Hosub, S. H. Rhim, J. Im, and A. Freeman, Scientific
Reports \textbf{3}, 1651 (2013).

\bibitem{Yan14} B. Yan, M. Jansen, and C. Felser, Nature Physics \textbf{9},
709 (2013).

\bibitem{Kul99} V. A. Kulbachinskii, N. Miura, H. Arimoto, T. Ikaida, P.
Lostak, and C. Drasar, J. Phys. Jpn. \textbf{68}, 3328 (1999).

\bibitem{Ste07} N. P. Stepanov, S. A. Nemov, M. K. Zhitinskaya, and T. E.
Svechinikova, Semiconductor \textbf{41}, 786 (2007).

\bibitem{But10} N. P. Butch, K. Kirshenbaum, P. Syers, A. B. Sushkov, G. S.
Jenkins, H. D. Drew, and J. Paglione, Phys. Rev. B \textbf{81}, 241301(R)
(2010).

\bibitem{Laf10} A. D. LaForge, A. Frenzel, B. C. Pursley, T. Lin, X. Liu, J. Shi,
and D. N. Basov, Phys. Rev. B \textbf{81}, 125120 (2010).

\bibitem{Sch12} A. A. Schafgans, K. W. Post, A. A. Taskin, Y. Ando, X.-L. Qi,
B. C. Chapler, and D. N. Basov, Phys. Rev. B \textbf{85}, 195440 (2012).

\bibitem{Pie12} P. Di Pietro, F. M. Vitucci, D. Nicoletti, L. Baldassarre, P.
Calvani, R. Cava, Y. S. Hor, U. Schade, and S. Lupi, Phys. Rev. B \textbf{86,}
045439 (2012).

\bibitem{Val12} R. Valdes Aguilar, A.V. Stier, W. Liu, L.S. Bilbro, D.K. George,
N. Bansal, L. Wu, J. Cerne, A.G. Markelz, S. Oh, and N. P. Armitage, Phys. Rev. Lett. \textbf{108},
087403 (2012).

\bibitem{Cha14} B. C. Chapler, K. W. Post, A. R. Richardella, J. S. Lee, J.
Tao, N. Samarth, and D. N. Basov, Phys. Rev. B \textbf{89}, 235308 (2014).

\bibitem{Wu15} L. Wu, W-K Tse, M. Brahlek, C.M. Morris, R. Valdes Aguilar,
N. Koirala, S. Oh, and N. P. Armitage, Phys. Rev. Lett. \textbf{115}, 217602-1 (2015)

\bibitem{Orl15} M. Orlita, B. A. Piot, G. Martinez, N. K. Sampath Kumar, C.
Faugeras, M. Potemski, C. Michel, E. M. Hankiewicz, T. Brauner, C. Drasar,
S. Schreyeck, S. Grauer, K. Brunner, C. Gould, C. Brune, and L. W. Molenkamp,
Phys. Rev. Lett. \textbf{114}, 186401 (2015).

\bibitem{Fan61} U. Fano, Phys. Rev. \textbf{124}, 1866 (1961).

\bibitem{Ric77} C. R. Richter, W. Kohler, and H. Becker, Phys. Stat. Sol. B 
\textbf{84}, 619 (1977).

\bibitem{Kul84} W. Kullmann, J. Geurts, W. Richter, H. Rauh, U.
Steigenbbrger, G. Eichhorn, and R. Geick, Phys. Stat. Sol. B \textbf{125},
131 (1984).

\bibitem{Wol64} P. A. Wolff, J. Phys. Chem. Solids\textbf{\ 25}, 1057 (1964).

\bibitem{Pal70} E. D. Palik and J. K. Furdyna, Rep. Prog. Phys. \textbf{33},
1193 (1970).

\bibitem{Bla57} E. M. Black, L. Conwell, L. Seigle, and C.vW. Spencer, J. Phys.
Chem. Solids \textbf{2}, 240 (1957).

\bibitem{Book} ``Bismuth telluride (Bi$_{2}$Te$_{3}$) effective masses", 
\textit{Non-Terrahedrally Bonded Elements and Binary Compounds I} of the
series \textit{Landolt-Bornstein-Group III Condendensed Matter}, Volume 41C
by Springer Berlin Heidelberg (1998).

\bibitem{Kul94} V. A. Kulbachinskii, M. Inoue, M. Sasaki, H. Negishi, W. X.
Gao, K. Takase, Y. Giman, P. Lostak and J. Horak, Phys. Rev. B \textbf{50},
16921 (1994).

\bibitem{Kuz05} A. B. Kuzmenko, Rev. Sci. Instrum. \textbf{76}, 093108 (2005).

\bibitem{Haj14} M. Hajlaoui, E. Papalazarou, J. Mauchain, L. Perfetti, A.
Taleb-Ibrahimi, F. Navarin, M. Monteverde, P. Auban-Senzier, C. R. Pasquier,
N. Moisan, D. Boschetto, M. Neupane, M. Z. Hasan, T. Durakiewicz, Z. Jiang,
Y. Xu, I. Miotkowski, Y. P. Chen, S. Jia, H. W. Ji, R. J. Cava, and M. Marsi,
Nature Communication \textbf{5}, 3003 (2014).

\bibitem{Ric92} M. J. Rice and H.-Y. Choi, Phys. Rev. B \textbf{45}, 10173
(1992).

\bibitem{Cap12} E. Cappelluti, L. Benfatto, M. Manzardo, and A. B. Kuzmenko,
Phys. Rev. B \textbf{86}, 115439 (2012).

\bibitem{Ess09} A. M. Essin, J. E. Moore, and D. Vanderbilt, Phys. Rev. Lett. 
\textbf{102}, 146805 (2009).

\bibitem{Koh76} H. Kohler, Phys. Status Solidi B \textbf{74}, 591 (1976).

\bibitem{Haj12} M. Hajlaoui, E. Papalazarou, J. Mauchain, G. Lantz, N.
Moisan, D. Boschetto, Z. Jiang, I. Miotkowski, Y. P. Chen, A.
Taleb-Ibrahimi, L. Perfetti, and M. Marsi, Nano Letters \textbf{12}, 3532
(2012).
\end{thebibliography}
\end{document}